\documentstyle[aps,prbbib,twocolumn,epsf]{revtex}

\begin{document}
\draft
\title{Nonlinear Spin Polarized Transport Through a Quantum Dot} 
\author{Baigeng Wang$^1$, Jian Wang $^1$ and Hong Guo$^2$}
\address{1. Department of Physics, The University of Hong Kong, \\
Pokfulam Road, Hong Kong, China\\
2. Center for the Physics of Materials and Department \\
of Physics, McGill University, Montreal, PQ, Canada H3A 2T8\\
}
\maketitle

\begin{abstract}
We present a theoretical analysis of the nonlinear bias and temperature
dependence of current-voltage characteristics of a spin-valve device
which is formed by connecting a quantum dot to two ferromagnetic electrodes
whose magnetic moments orient at an angle $\theta$ with respect to each 
other. 
The theory is based on nonequilibrium Green's function approach and focused
on current perpendicular to plane geometry. Coulomb interaction has been 
taken into account explicitly at the Hartree level. We derive a formula in 
closed form for current flowing through the device in general terms of bias 
and temperature. In the wideband limit we report exact results for the 
TMR junction nonlinear I-V curve as a function of $\theta$. We also report 
the conductance slope at zero bias as a function of temperature for which
experimental results reported an anomalous behavior.
\end{abstract}


\section{Introduction}

Due to the exciting perspective of magnetoelectronics where coherent charge
and polarized spin are both utilized for electronic device 
functions\cite{prinz}, quantum transport properties of systems with a 
magnetic character have attracted extensive attention. The well known giant 
magnetoresistance effect (GMR) is such a spin-polarized electronic 
transport effect\cite{baibich,meservey}, using which a GMR device changes its
resistance depending on the orientation of magnetic moments of the magnetic
multi-layers forming the device. A GMR system is fabricated by sandwiching
non-magnetic metal layer between two magnetic layers\cite{baibich}, and 
further extensions or variations of this structure can produce many other 
device functions, including the spin-valve transistor\cite{monsma}, 
spin selective electron interferometer\cite{egger}, and nonvolatile 
RAM\cite{dax}. While GMR is perhaps one of the most fruitful areas of
fundamental research which has already produced substantial application with
huge commercial value, an entirely different approach, also exploiting 
spin-polarized transport, is the tunneling magnetoresistance (TMR) where an 
insulating (I) material layer is sandwiched in between two ferromagnetic 
(FM) layers\cite{jull}, forming a $FM/I/FM$ structure. One important 
attraction of a TMR device is that it carries lower current than the metallic 
GMR system, hence TMR can be, in principle, more advantageous for portable
devices. Due to the reported room temperature operation of TMR\cite{moodera}, 
the fundamental principle and various transport characteristics of TMR 
devices have been the subject of increasingly more research, and will be the
topic of the present work.

Tunneling of spin-polarized electrons through the TMR structure gives 
a large tunneling magnetoresistance (TMR) through the spin-valve 
effect\cite{jull,slon}. The spin valve effect is characterized by the 
fractional change of resistance $\kappa \equiv \Delta R/R$, where $\Delta R$ 
is the difference in resistance when magnetization of the two ferromagnetic 
metals are in parallel or anti-parallel. In the original experiment 
of Julliere on $Co/Ge/Fe$, $\kappa$ was determined to be about 
$14\%$ at $4.2$K temperature at zero external bias voltage, and it 
dropped to $1\%$ when bias was increased to a few meV. More recently,
large TMR was reported\cite{moodera,moodera1} to persist at room temperature
where $\kappa$ is about $11.8\%$ for $CoFe/Al_2O_3/Co$ at small bias. 
Again, in this device $\kappa$ decreases drastically when the voltage 
increases\cite{moodera,moodera1}. These experimental results clearly point 
to the need of a thorough theoretical understanding of the bias as well as
temperature dependence of TMR. On the theoretical side, TMR has been
investigated by many authors. Zhang {\it et.al.} have considered\cite{zhang}
TMR using a transfer matrix and calculated the tunnel conductance and
magnetoresistance of a FM/I/FM system with an oscillatory but decaying 
bias voltage dependence of TMR resulting from quantum resonance.
Bratkovsky\cite{bratkovsky} investigated a similar system and compared cases
with or without impurity scattering. Barna\'s and Fert investigated  
magnetoresistance oscillations due to charging effects\cite{fert}; Zhang
{\it et.al.}\cite{levy} investigated the zero-bias anomaly of the 
experimental data\cite{moodera} and provided an explanation that the 
anomaly is 
due to localized spin excitations at the barrier-electrode interface; and a
number of studies\cite{zhang,bratkovsky,inoue,maclaren,mathon}
were devoted to improve the original TMR model of Julliere\cite{jull}.
Very recently, the nonlinear bias dependence of an interesting TMR 
device in the form of $FM/I/FM/I/FM$ was investigated 
theoretically\cite{sheng}. Due to the relative magnetic moment orientation 
of the three FM layers as well as the quantum well formed by the two 
insulating layers, various very interesting TMR behavior can be 
expected\cite{sheng}.

The purpose of this work is to further investigate TMR effect as a function
of external bias as well as temperature in a $FM/I/FM$ device structure,
where we pay special attention to the internal Coulomb potential due to
interactions. For a quantum device subjected to an external bias voltage, 
the internal potential build-up can be an important factor determining 
transport characteristics at the nonlinear regime\cite{but1}, and this has 
not been included in previous analysis 
Ref.\onlinecite{zhang,bratkovsky,sheng,levy}
of the bias dependence of TMR. There, external bias potentials are
included non-self-consistently through Fermi functions of the electron
reservoirs. In this paper we present a theory for TMR junctions based on 
the nonequilibrium Green's function, where the internal potential is solved 
self-consistently at the Hartree level. In particular we derive a general 
expression for the electric current flowing through the TMR device in terms 
of external bias, temperature, and the angle $\theta$ between the magnetic 
moments of the two FM electrodes.  Due to the inclusion of the internal 
Coulomb potential into the theory, the predicted results are gauge invariant 
against the shift of the external bias --- which is a fundamental 
requirement of any transport theory. In the wideband limit
and applying the charging model for the internal potential, we further derive 
an analytic expression of angular dependence of the I-V curve at zero 
temperature. The temperature dependence of TMR is then studied via a 
numerical integration of our analytical formula. Finally, in the weakly
nonlinear regime our theory allows predictions of the slope of the TMR near
zero bias, where experiments\cite{moodera} showed a dip of magneto-conductance
at low temperatures. Our results showed that the slope smoothly crosses over 
from a finite value at low temperature to zero at high temperature.

The rest of this paper is organized as follows. In section II, we present 
the general theory by deriving a current expression for TMR junctions. 
In section III, we apply this current expression for various situations, 
and a summary is given in section IV.

\section{General theoretical formulation}

The TMR device we examine is schematically shown in the inset of Fig. (1a). 
It consists of a quantum dot connected by two ferromagnetic electrodes to 
the outside
world. The magnetic moment ${\bf M}$ of the left electrode is pointing to 
the $z$-direction, the electric current is flowing in the $y$-direction, 
while the moment of the right electrode is at an angle $\theta$ to the 
$z$-axis in the $x-z$ plane. In second quantized form this device is
described by the following Hamiltonian,
\begin{equation}
H = H_L+H_R+H_{dot}+H_T
\label{eq1}
\end{equation}
where $H_L$ and $H_R$ describe the left and right electrodes 
where a DC bias potential is applied,
\begin{equation}
H_L = \sum_{k\sigma} (\epsilon_{kL}+\sigma M) ~ c^{\dagger}_{kL\sigma}
c_{kL\sigma}
\label{eq2}
\end{equation}
\begin{eqnarray}
H_R &=& \sum_{k\sigma} [(\epsilon_{kR}+\sigma M\cos\theta) ~
c^{\dagger}_{kR\sigma} c_{kR\sigma}  \nonumber \\
&+& \sum_{k\sigma} M\sin\theta ~ [c^{\dagger}_{kR\sigma} c_{kR\bar{\sigma}} ]
\ \ .
\label{eq3}
\end{eqnarray}
In Eq. (\ref{eq1}), $H_{dot}$ describes the quantum dot
\begin{equation}
H_{dot}= \sum_{n\sigma} \epsilon_n d^{\dagger}_{n\sigma} d_{n\sigma}\ \ .
\label{hdot}
\end{equation}
$H_T$ models the coupling between electrodes and the quantum dot region (the 
scattering region) with hopping matrix $T_{k\alpha n}$. For simplicity we 
shall assume the hopping matrix to be independent of spin index,  thus
\begin{equation}
H_T= \sum_{k \alpha n\sigma} [T_{k\alpha n} ~ 
c^{\dagger}_{k\alpha\sigma} d_{n\sigma} + c.c.]\ \ .
\label{ht}
\end{equation}
In these expressions $\epsilon_{k\alpha}= \epsilon^0_{k} + q {V}_{\alpha}$ 
with $\alpha=L,R$; $c^{\dagger}_{k\alpha\sigma}$ (with $\sigma=\uparrow,
\downarrow$ or $\pm 1$ and $\bar{\sigma}=-\sigma$) is the creation operator 
of electrons with spin index $\sigma$ inside the $\alpha$-electrode. 
Similarly $d^{\dagger}_{n\sigma}$ is the creation operator of electrons 
with spin $\sigma$ at energy level $n$ for the quantum dot region. In writing 
down Eqs.(\ref{eq2}) and (\ref{eq3}), we have made a simplification
that the value of molecular field $M$ is the same for the two electrodes, 
thus the spin-valve effect is obtained\cite{slon} by varying the angle 
$\theta$. Essentially, $M$ mimics the difference of density of states 
(DOS) between spin up and down electrons\cite{slon} in the electrodes. 
Finally, due to electron-electron interactions an internal potential build-up 
$U({\bf r})$ is induced inside the quantum dot region. Hence the actual 
Hamiltonian of the quantum dot is $H_{dot}+qU$. 

To proceed, we first apply the following Bogliubov transformation\cite{bog} 
to diagonalize the Hamiltonian of the right electrode, 
\begin{equation}
c_{kR\sigma }\rightarrow \cos (\theta/2)C_{kR\sigma }-\sigma \sin (
\theta/2)C_{kR\bar{\sigma}}
\end{equation}
\begin{equation}
c_{kR\sigma }^{+}\rightarrow \cos (\theta/2)C_{kR\sigma
}^{+}-\sigma \sin (\theta/2)C_{kR\bar{\sigma}}^{+}
\end{equation}
from which we obtain the effective Hamiltonian
\begin{equation}
H_\alpha = \sum_{k\sigma} [(\epsilon_{k\alpha}+\sigma M)
C^{\dagger}_{k\alpha\sigma}C_{k\alpha\sigma}
\label{halpha}
\end{equation}
\begin{eqnarray}
H_T &=& \sum_{k n\sigma} [T_{kLn\sigma} ~ C^{\dagger}_{kL\sigma} 
d_{n\sigma} + T_{kRn\sigma} (\cos\frac{\theta}{2} ~ C^{\dagger}_{kR\sigma}  
\nonumber \\
&-& \sigma \sin\frac{\theta}{2} ~ C^{\dagger}_{kR\bar{\sigma}}) ~ 
d_{n\sigma} +c.c.]\ \ .
\label{eff}
\end{eqnarray}

The electric current flowing the TMR device modeled by Eqs. (\ref{eq1},
\ref{halpha}, \ref{hdot}, \ref{eff}) can be written, following the
derivations of Ref. \onlinecite{jauho} and extending it to include
spin dependent scattering, in terms of the Green's functions of the 
quantum dot ($\hbar =1$), 
\begin{eqnarray}
I_\alpha &=& iq \int (dE/\pi) Tr ~ \left\{ -2Im({\bf \Sigma}^r_{\alpha} ) 
\right.  \nonumber \\
& \times & \left. \left[({\bf G}^r-{\bf G}^a) f_\alpha +{\bf G}^< 
\right]\right\}
\label{i1}
\end{eqnarray}
where $f_\alpha \equiv f(E-qV_\alpha)$ and the trace is over both the
state index and spin index. Here ${\bf G}^r(E,U)$ is the $2\times 2$ 
matrix in spin space for the retarded Green's function with $U({\bf r})$ 
the electro-static potential build-up inside quantum dot region. 
In Hartree approximation ${\bf G}^r(E,U)$ is given by
\begin{equation}
{\bf G}^r(E,U) = \frac{1}{E-H_{dot} - qU -{\bf \Sigma}^r}  \label{gr}
\end{equation}
where the self-energy ${\bf \Sigma}^r\equiv {\bf \Sigma}^r_L(E-qV_L) + 
{\bf \Sigma}^r_R(E-qV_R)$ is also a $2\times2$ matrix in spin space, which
describes coupling of the quantum dot region to the two magnetic electrodes.
The detailed analysis of self-energy is given in the Appendix and 
the result is written as
\begin{equation}
{\bf \Sigma}^r_\alpha(E) = \hat{R}_\alpha \left( 
\begin{array}{cc}
\Sigma^r_{\alpha\uparrow} & 0 \\ 
0 & \Sigma^r_{\alpha\downarrow}
\end{array}
\right) \hat{R}^\dagger_\alpha
\label{self}
\end{equation}
with the rotational matrix $\hat{R}_\alpha$ for electrode $\alpha$ 
defined as 
\begin{equation}
\hat{R} = \left( 
\begin{array}{cc}
\cos\theta_\alpha/2 ~~ & \sin\theta_\alpha/2 \\ 
-\sin\theta_\alpha/2 ~~ & \cos\theta_\alpha/2
\end{array}
\right)\ \ .
\end{equation}
Here angle $\theta_{\alpha}$ is defined as $\theta_L=0$ and $\theta_R=\theta$ 
and $\Sigma^r_{\alpha \sigma}$ is given by 
\begin{equation}
\Sigma^r_{\alpha \sigma mn}= \sum_k \frac{T^*_{k\alpha m} T_{k\alpha n}} 
{E-\epsilon_{k\alpha \sigma} +i\delta}
\end{equation}
The lesser Green's function ${\bf G}^{<}$ in Eq. (\ref{i1}) is 
calculated\cite{jauho} through the Keldysh equation 
${\bf G}^{<}={\bf G}^{r}{\bf \Sigma }^{<}{\bf G}^{a}$ with the 
self energy ${\bf \Sigma }^{<}={\bf \Sigma }_{L}^{<}(E-qV_{L})+{\bf \Sigma}
_{R}^{<}(E-qV_{R})$ given by the following expression at equilibrium,
\begin{equation}
{\bf \Sigma}^<_\alpha(E) = i f_\alpha \hat{R}_\alpha \left( 
\begin{array}{cc}
\Gamma_{\alpha\uparrow} & 0 \\ 
0 & \Gamma_{\alpha\downarrow}
\end{array}
\right) \hat{R}_\alpha^\dagger
\label{ss}
\end{equation}
where $\Gamma_{\alpha\sigma}= -2Im(\Sigma^r_{\alpha\sigma})$ is the
linewidth function. 

Using Keldysh equation and the explicit self-energy expressions 
(\ref{self},\ref{ss}), after tedious but straightforward manipulations 
one can confirm that Eq.(\ref{i1}) becomes
\begin{equation}
I_\alpha =\frac{2q}{\pi}\int dE \sum_{\beta\neq \alpha} 
Tr \left[ Im({\bf \Sigma}^r_\alpha) {\bf G}^r Im({\bf \Sigma}^r_\beta) 
{\bf G}^a \right] (f_\alpha - f_\beta)\ .
\label{final1}
\end{equation}
Note that the same form of current expression is true for systems without 
spin dependent scattering\cite{datta1}. But with such scattering, the trace 
is over states as well as spin variables. Eq. (\ref{final1}) is the basis 
for our further analysis.

Before presenting the results for the Green's function ${\bf G}^r$, 
it is worth to emphasize that it is now explicitly dependent on the
internal potential landscape through the Hartree potential $U({\bf r})$. 
This is an extension to the previous NEGF analysis\cite{jauho,datta,stafford} 
and it allows us to investigate TMR within gauge invariance requirement. 
At Hartree level $U({\bf r})$ is determined by the self-consistent Poisson 
equation 
\begin{equation}
\nabla^2 U = 4\pi i q\int (dE/2\pi)\sum_\sigma ({\bf G}^<(E,U))_{\sigma
\sigma}  \ \ .
\label{poisson}
\end{equation}
Eq. (\ref{poisson}) is a {\it nonlinear} equation because 
${\bf G}^{r,a}$ depends on $U({\bf r})$ by Eq.(\ref{gr}). Hence in this
theory there is a need to self-consistently solve the coupled equations
of ${\bf G}^{r,a}$ and $U({\bf r})$. This presents some analytical
difficulties (see next section).

Eqs. (\ref{final1},\ref{gr},\ref{poisson}) completely determine the
nonlinear I-V characteristics of multi-probe TMR junctions. It is not 
difficult to directly prove that the current expression Eq.(\ref{final1}) 
is gauge invariant: shifting the potential at all the electrodes by a 
constant $\Delta V$, 
$V_{\alpha}\rightarrow V_{\alpha}+\Delta V$, $I_\alpha$ from 
Eq. (\ref{final1}) remains the same.  

From now on we focus on the two-probe TMR device of the inset of 
Fig. (1a) for which $I_L=-I_R=I$. After expanding the spin part of the 
trace in Eq. (\ref{final1}), we obtain the final expression of electric
current for the TMR device,
\begin{eqnarray}
& & I = \frac{q}{2\pi} \int dE Tr\left[ \right. \nonumber \\
& & \left. \Gamma_{L\uparrow} G^r_{11} (\Gamma_{R\uparrow} 
\cos^2 \frac{\theta}{2} + \Gamma_{R\downarrow} \sin^2 
\frac{\theta}{2} ) G^a_{11} \right.  \nonumber \\
& &-\Gamma_{L\uparrow} G^r_{12} (2\Gamma_{L\downarrow} + 
\Gamma_{R\downarrow} \cos^2 \frac{\theta}{2} + \Gamma_{R\uparrow} \sin^2 
\frac{\theta}{2}) G^a_{12} \nonumber \\
& &-\Gamma_{L\downarrow} G^r_{21} (2\Gamma_{L\uparrow} + 
\Gamma_{R\uparrow} \cos^2 \frac{\theta}{2} + \Gamma_{R\downarrow} 
\sin^2 \frac{\theta}{2}) G^a_{21} 
\nonumber \\
& &+\left.\Gamma_{L\downarrow} G^r_{22} (\Gamma_{R\downarrow} 
\cos^2\frac{\theta}{2} 
+ \Gamma_{R\uparrow} \sin^2 \frac{\theta}{2} ) G^a_{22}  
\right](f_R - f_L)\ .
\label{current}
\end{eqnarray}
In this result, the $2\times 2$ Green's function matrix ${\bf G}^r$
is determined by the usual method of equation of motion\cite{jauho}. For
our Hamiltonian the equation of motion can be solved exactly through
straightforward algebra, and they are found to be:
\begin{equation}
G^r_{11}\equiv G^r_{\uparrow\uparrow} 
= \frac{A_\downarrow}{A_\uparrow A_\downarrow -B^2}
\label{g11}
\end{equation}
\begin{equation}
G^r_{12}\equiv G^r_{\uparrow\downarrow} 
= G^r_{21}\equiv G^r_{\downarrow\uparrow} 
= \frac{B}{A_\uparrow A_\downarrow -B^2}
\label{g12}
\end{equation}
\begin{equation}
G^r_{22}\equiv G^r_{\downarrow\downarrow}
= \frac{A_\uparrow}{A_\uparrow A_\downarrow -B^2}
\label{g22}
\end{equation}
where $A_\sigma$ and $B$ are defined as
\begin{equation}
A_\sigma = E -H_{dot} -qU -\Sigma^r_\sigma
\end{equation}
\begin{equation}
B = \frac{i}{2} (\Gamma_{R\downarrow} - \Gamma_{R\uparrow}) 
\sin\frac{\theta}{2} \cos\frac{\theta}{2}\ \ .
\end{equation}

Before finishing this section, we note that Eqs. (\ref{current}, \ref{g11}, 
\ref{g12}, \ref{g22}, \ref{poisson}) form the basis for numerical 
calculations of TMR I-V curves beyond what has been done before, since it is
a gauge invariant formula. In a typical numerical analysis, one
computes Green's functions ${\bf G}^r$ and the coupling matrix $\Gamma$ 
using tight-binding models\cite{datta1}; and the Poisson equation can be 
solved using very powerful numerical techniques\cite{jwang}.
Numerical analysis of these expressions is beyond the scope of this paper
and in the next section we attempt to obtain the I-V characteristics via
analytic means.

\section{Applications}

As discussed in the last section, the spin dependent nonlinear I-V 
characteristics of the TMR device is obtained by self-consistently
solving Eqs. (\ref{current},\ref{g11}, \ref{g12}, \ref{g22}, \ref{poisson}). 
Due to complexity of this problem, some reasonable approximations need to
be made. In particular we will present applications of our theory
from two approximations. First, instead of solving the Poisson equation 
(\ref{poisson}) for $U({\bf r})$ analytically which is perhaps 
impossible to do, we will parameterize $U$ in terms of related geometrical 
capacitance. This is the discrete potential approximation commonly used 
in theoretical analysis\cite{christen,stafford}. This analysis allows 
us to calculate general nonlinear I-V curves at various temperature and 
qualitatively compare with the experimental observations. Second, at weakly 
nonlinear regime, the bias voltage is finite but small hence one can expand 
the entire current formula in terms of bias order by order. In this case the 
Poisson equation can be directly solved to yield a spatially varying 
$U({\bf r})$. At weakly nonlinear regime, we calculate the second order 
nonlinear conductance which is the slope of the I-V curve at zero bias.
Experimentally\cite{moodera} one observes a finite slope at low temperature,
while zero slope at high temperature.

\subsection{TMR I-V curves and ratios}

To obtain a general qualitative behavior of the I-V curves, 
we shall simplify the current
formula (\ref{current}) using the discrete potential approximation. In this 
approximation\cite{christen,pedersen} one assumes that $U({\bf r})=U_o$ 
is a constant that provides a shift to the band bottom of the quantum dot 
due to Coulomb interactions. To solve for $U_o$, we introduce geometrical
capacitance coefficients $C_1$ and $C_2$ between the electrodes and the 
quantum dot. The total charge $\Delta Q$ in the quantum dot is given by the
right hand side of Eq. (\ref{poisson}), and we parameterize $\Delta Q$
using $C_1$ and $C_2$,
\begin{eqnarray}
\Delta Q&=&-\frac{i}{2\pi}\int dE Tr\left[{\bf G}^<(E,U_0)-
{\bf G}^<_0\right] \nonumber \\
&=& C_1\times (U_0-V_1) + C_2\times (U_0-V_2)\ \ .
\label{ex}
\end{eqnarray}
Hence by evaluating the energy integration and using $C_1$, $C_2$ as
input parameters, one obtains $U_o$.

The explicit expressions for the Green's functions and hence the current
(\ref{current}) and charge (\ref{ex}), can be derived in the wideband 
limit\cite{jauho} where the coupling matrix $\Gamma$ is independent 
of energy. In this limit the energy integrations of (\ref{current},
\ref{ex}) can be completed. For instance, assuming $\Gamma_{L\sigma}
=\Gamma_{R\sigma}\equiv \Gamma_\sigma$, at zero temperature we obtain the 
I-V curve of the TMR device to be,
\begin{eqnarray}
I &=&  \sum_\sigma p_\sigma \left(\arctan\frac{2E_1}{\Gamma+\sigma \Delta 
\Gamma \cos\theta/2} \right.\nonumber \\
&-& \left. \arctan\frac{2E_2}{\Gamma+\sigma \Delta \Gamma 
\cos\theta/2} \right)
\label{current1}
\end{eqnarray}
where $\Gamma=\Gamma_\uparrow+\Gamma_\downarrow$, $\Delta 
\Gamma=\Gamma_\uparrow-\Gamma_\downarrow$, $E_\alpha = E_F - E_0 -qU_0
+qV_\alpha$, and 
\[p_\sigma = -\frac{q}{4\pi \Gamma} (\Gamma^2 + \sigma \Gamma \Delta \Gamma
\cos\frac{\theta}{2} - \Delta \Gamma^2 \sin^2\frac{\theta}{2})\ .
\]
The internal potential $U_o$ is determined by the following equation
obtained by evaluating the charge $\Delta Q$ at the wideband limit,
\begin{eqnarray}
& & \sum_{\alpha\sigma} \left[ \arctan\frac{2E_\alpha}{\Gamma+\sigma \Delta 
\Gamma \cos\theta/2} \right. \nonumber \\
&-& \left. \arctan\frac{2(E_F-E_0)}{\Gamma+\sigma \Delta \Gamma 
\cos\theta/2} \right] \nonumber \\
&=& \frac{2\pi}{q}[C_1 (U_0-V_1)+C_2 (U_0-V_2)]\ .
\label{potential1}
\end{eqnarray}
To get physical insight we plot the I-V characteristics 
at different temperature parameter $\beta$ and magnetic moment orientation
$\theta$. We have chosen $\Gamma_\uparrow=1.0$, $\Gamma_\downarrow=0.4$, 
$C_1=C_2=0.5$ and $E_F-E_0=-2$ (units set by $\hbar=e=2m=1$). 

Fig.(1a) shows I-V curves at zero temperature $\beta=\infty$ for different 
orientations $\theta=0$ (solid line), $0.4\pi$ (dotted line), 
and $0.8\pi$ (dashed line). Clearly current increases as the junction bias
increases. When $\theta=0$, {\it i.e.} when magnetic moment of the left 
and right electrodes are parallel, the current is largest at all bias
voltages. When $\theta=\pi$ for which the moments are anti-parallel, the
current is the smallest. This is a well known result for TMR junctions at
zero bias\cite{slon}, but it holds at finite bias as well. This behavior 
can be seen more clearly in Fig.(1b) where we plot current versus angle
$\theta$ at different bias $\Delta V=(V_1-V_2)=2, 5, 8, 11, 14$. 
For all cases $I$ is minimum at $\theta=\pi$.
When temperature is nonzero the energy integration of Eq. (\ref{current})
cannot be completed analytically thus we integrate it numerically assuming
phonons can be neglected. Fig. (2) plots current versus $\theta$ at 
different temperatures $\beta=1,0.5,0.2$. The current is sensitively
dependent on temperature and decreases substantially when temperature is
increased (smaller $\beta$ corresponds to higher temperature). This is 
consistent with previous experimental findings\cite{jull,moodera}.

To calculate tunneling magnetoresistance ratio, we use the definition 
of Ref. \onlinecite{sheng} which defines a nonlinear resistance 
$R=1/G\equiv V/I$ and TMR ratio $\equiv (R(0)-R(\pi))/R(0)$. The TMR 
ratio versus voltage is presented in Fig.(3) where two curves corresponding 
to $\beta=\infty$ and $\beta =1$ are plotted respectively. At zero 
temperature, a large TMR ratio (about $30\%$) drops quickly as the voltage 
is increased, a trend agrees with experimental findings\cite{jull,moodera}. 
At small voltages, the TMR ratio is much lower for higher temperature, also
in agreement with experimental results\cite{moodera}. 
We note that the theoretical values of the TMR ratio, about $34\%$ at low
temperature, are much larger than the measured values\cite{moodera1} 
(about $27\%$ at $77$K), due to our choice of ideal model and system 
parameters. But the qualitative trend of these results are consistent.
The TMR minimum in Fig. (3) is due to quantum resonance, as was also seen in
previous studies\cite{zhang}. Finally, we note that experimental data of TMR
as a function of bias is a slight concave-shaped curve near zero bias,
but that of our model, as well as of others\cite{bratkovsky,zhang,sheng}, 
are slightly convex. Thus the decay rate of TMR is faster in the measured
data\cite{moodera1}, which indicates that other physical 
mechanisms\cite{levy} absent in the present model are perhaps needed to 
quantitatively understand the experimental data.

\subsection{Conductance dip at zero bias}

Experimental results on TMR junctions of Ref. \onlinecite{moodera} showed 
that there is a clear conductance dip at zero bias at low temperature,
a slight dip at intermediate temperature, and perhaps no dip at high 
temperature. This anomaly may be associated with a number of device details 
such as the presence of impurities, localization, and scattering. In this 
section we will examine the zero-bias conductance dip as a function of
temperature.

The slope of conductance versus bias is the second order nonlinear 
conductance coefficient $G_{\alpha \beta \gamma}
\equiv d^2I_\alpha/dV_\beta dV_\gamma$ evaluated at zero bias voltages
$V_\beta\rightarrow 0$, $V_\gamma\rightarrow 0$. We apply a weakly 
nonlinear analysis of $G_{\alpha \beta \gamma}$ in which all quantities 
of interest are expanded order by order in bias voltage. This way one can
directly solve the Poisson equation (\ref{poisson}) without using the
discrete potential approximation:  we seek the solution of $U({\bf r})$ 
in the following form,
\begin{equation}
U= U_{eq} + \sum_{\alpha} u_{\alpha} V_{\alpha} +\frac{1}{2}\sum_{\alpha 
\beta} u_{\alpha \beta} V_{\alpha} V_{\beta} + ...
\label{char}
\end{equation}
where $U_{eq}$ is the equilibrium potential and $u_\alpha({\bf r})$, 
$u_{\alpha \beta}({\bf r})$ are the characteristic potentials\cite{but1,ma1}. 
It can be shown that the characteristic potential satisfy the following 
sum rules\cite{but1,ma1} $\sum_{\alpha} u_{\alpha} =1$ and
$\sum_{\beta} u_{\alpha\beta}=0$. Expanding ${\bf G}^<$ of Eq. (\ref{poisson}) 
in power series of $V_\alpha$, from the Poisson equation (\ref{poisson}) 
one can derive equations for all the characteristic potentials. In particular 
the expansions are facilitated by Dyson equation to the appropriate 
order\cite{wbg1}:  
\[
{\bf G}^r = {\bf G}^r_0 + {\bf G}^r_0 (qU-qU_{eq}) {\bf G}^r_0 +\cdots 
\]
with ${\bf G}^r_0$ the equilibrium retarded Green's function, {\it i.e.}, 
when $U=U_{eq}$. At the lowest order we obtain\cite{foot}
\begin{eqnarray}
-\nabla^2 u_{\alpha}({\bf r}) 
&=& -4\pi q^2 \frac{dn({\bf r})}{dE} u_\alpha({\bf r})
\nonumber \\
&+& 4\pi q^2 \frac{dn_{\alpha}({\bf r})}{dE}
\label{x1}
\end{eqnarray}
The first term on the right hand side of Eq. (\ref{x1}), which depends
on internal potential $u_\alpha$, describes the induced charge density in 
the TMR junction. The second term of Eq.(\ref{x1}) is the {\it injectivity} 
which corresponds to the charge density due to external injection.  Finally
\begin{equation}
\frac{dn_{\alpha}(x)}{dE} =  -2\sum_\sigma \int \frac{dE}{2\pi} 
(-\partial_E f)
[{\bf G}^r_0 Im{\bf \Sigma}^r_{\alpha} {\bf G}^a_0]_{\sigma \sigma} 
\label{inj} 
\end{equation}
and 
\begin{equation}
\frac{dn}{dE} = \sum_{\alpha} \frac{dn_{\alpha}}{dE} 
\end{equation}
are the local density of states. Once the characteristic potential is 
obtained from Eq.(\ref{x1}), the the second order nonlinear spin dependent 
conductance $G_{\alpha \beta \gamma}$ is given by
\begin{eqnarray}
& & G_{\alpha \beta \gamma} = -\frac{2q^3}{2\pi}\int dE
(-\partial_E f) Tr\left[ 
({\bf G}^a_0 Im{\bf \Sigma}^r_{\alpha} {\bf G}^r_0 {\bf G}^r_0 
\right.
\nonumber \\
&+& \left. {\bf G}^a_0 {\bf G}^a_0 Im{\bf \Sigma}^r_{\alpha} {\bf G}^r_0) 
(Im {\bf \Sigma}^r \delta_{\alpha \gamma} - Im{\bf \Sigma}^r_{\gamma}) 
(2u_\beta- \delta_{\beta \gamma})
\right]
\label{g111}
\end{eqnarray}
where $u_\beta$ in Eq.(\ref{g111}) is a $2 \times 2$ diagonal matrix.

The conductance slope at zero bias which has been measured in the 
experiments of Ref. \onlinecite{moodera,moodera1} is given by $G_{111}$, 
since $G_{111}$ is the slope of $dI/dV$ at zero bias. To simplify 
discussion we consider $G_{111}$ near a resonant point where the Green's 
function is given by $G^r_{0,11}=1/(E_F-E_0+i \Gamma_\uparrow/2)$ and 
$G^r_{0,22}=1/(E_F-E_0 +i \Gamma_\downarrow/2)$. In addition, apply 
the quasi-neutrality approximation\cite{but5} we have 
$u_1=(dn_1/dE)/(dn/dE)$. Thus expression (\ref{g111}) is easily calculated. 
In Fig. (4) we plot $G_{111}$ as a function of temperature at $E_F=2.5$ while
$E_0=3.0$. As the temperature is increased ($\beta$ decreases), 
$G_{111}$ goes to zero smoothly. Our model thus indicates that conductance 
slope $G_{111}$ is nonzero at small temperature, and it diminishes to zero 
at high temperatures, consistent with experimental results\cite{moodera}. 
Our result also indicates that the crossover from a finite slope to zero 
slope as a function of temperature is smooth, as shown in Fig. (4). 

\section{Summary}

In this paper, quantum transport properties of a mesoscopic conductor 
connected to two metallic ferromagnetic electrodes have been studied
theoretically. We derived self-consistent equations which completely 
determine nonlinear I-V characteristics of the TMR device. Our theory 
is gauge invariant due to the inclusion of long-range Coulomb potential. 
In the wideband limit the nonlinear I-V curves and the TMR ratio at 
different temperature and magnetic moment orientation $\theta$ are
calculated and our results are qualitatively consistent with the
experimental measurements. We were also able to investigate the conductance
slopes at zero bias for various temperatures, where experimental
measurements showed an anomaly. Experimental measurements\cite{moodera} 
gave data for a few temperatures, and our result shows that it is a smooth
crossover from a finite slope to zero slope as temperature is increased.
Finally, in this paper we concentrated on analytical predictions where
several commonly used approximations were applied. Our results indicate
that to completely explain all the anomalies in measured data of TMR, 
some further physical effects, which are neglected so far, may have to be 
included. 

\bigskip 
{\bf Acknowledgments.} We gratefully acknowledge support by a RGC grant 
from the SAR Government of Hong Kong under grant number HKU 7115/98P,
and a CRCG grant from the University of Hong Kong. H. G is supported by
NSERC of Canada and FCAR of Qu\'ebec. We thank the computer center of HKU
for computational facilities.

\section{Appendix}
In this appendix, we will derive the self-energy ${\bf \Sigma}^r_\alpha$.
For $\alpha =L,$ the self-energy can be written rather easily\cite{jauho}

\[
(\Sigma _{L}^{r})_{mn,\sigma \sigma ^{\prime }}=\sum_{k}T_{kLm}^{\ast
}T_{kLn}g_{kL\sigma }^{r}\delta _{\sigma \sigma ^{\prime }} 
\]
where $g^r_{k\alpha\sigma}$ is the free retarded Green's function for the
lead $\alpha$. 
However, for $\alpha =R,$ the situation becomes slightly complicated due to
the spin-flip process. In the following, we will use the Dyson equation to
calculate the self-energy function\cite{caroli}. To facilitate the 
calculation, we will define the following retarded Green's functions,

\begin{equation}
G_{mn,\sigma \sigma ^{\prime }}^{r}(t_{1},t_{2})\equiv -i\theta
(t_{1}-t_{2})\langle \{d_{m\sigma }(t_{1}),d_{n\sigma ^{\prime
}}^{+}(t_{2})\}\rangle
\end{equation}

\begin{equation}
G_{kn,\sigma \sigma ^{\prime }}^{r}(t_{1},t_{2})\equiv -i\theta
(t_{1}-t_{2})\langle \{C_{kR\sigma }(t_{1}),d_{n\sigma ^{\prime
}}^{+}(t_{2})\}\rangle
\end{equation}
where $m$ and $n$ label the states in the quantum dot and $k$ labels the
states in the lead. To illustrate the calculation procedure, we consider 
$G_{mn,\uparrow \uparrow }^{r}$ for simplicity. By using the Dyson equation

\begin{equation}
G^{r}=G_{0}^{r}+G_{0}^{r}\Sigma ^{r}G^{r}
\end{equation}
we obtain

\begin{eqnarray}
G^{r}_{mn,\uparrow \uparrow } &=&(G_{0}^{r})_{mn,\uparrow \uparrow
}+\sum_{k,m'}(G_{0}^{r})_{mm^{\prime },\uparrow \uparrow }\Sigma
^{r}_{m^{\prime }k,\uparrow \uparrow }G^{r}_{kn,\uparrow \uparrow }
\nonumber \\
&+&\sum_{k}(G_{0}^{r})_{mm^{\prime },\uparrow \uparrow }\Sigma
^{r}_{m^{\prime }k,\uparrow \downarrow }G^{r}_{kn,\downarrow \uparrow }
\label{ae1}
\end{eqnarray}
Using the Dyson equation again for the retarded Green's functions 
$G^{r}_{kn,\uparrow \uparrow }$ and $G^{r}_{kn,\downarrow \uparrow }$,
we have the following relations

\begin{equation}
G^{r}_{kn,\uparrow \uparrow }=g^r_{kR\uparrow }\Sigma ^{r}_{km,\uparrow
\uparrow }G^{r}_{mn,\uparrow \uparrow }+g^r_{kR\uparrow }\Sigma
^{r}_{km,\uparrow \downarrow }G^{r}_{mn,\downarrow \uparrow }
\label{ae2}
\end{equation}

\begin{equation}
G^{r}_{kn,\downarrow \uparrow }=g^r_{kR\downarrow }\Sigma
^{r}_{km,\downarrow \uparrow }G^{r}_{mn,\uparrow \uparrow
}+g^r_{kR\downarrow }\Sigma ^{r}_{km,\downarrow \downarrow
}G^{r}_{mn,\downarrow \uparrow }
\label{ae3}
\end{equation}
The self-energy matrix $\Sigma^r$ has the following matrix elements from 
Eq.(\ref{eff}): $(\Sigma ^{r})_{m^{\prime }k,\uparrow \uparrow }=\cos 
(\theta /2) T_{kRm^{\prime }}^{\ast }$ , $(\Sigma ^{r})_{m^{\prime }k,
\uparrow \downarrow }$ $=-\sin (\theta /2)T_{kRm^{\prime }}^{\ast }$ ,
$(\Sigma ^{r})_{m^{\prime }k,\downarrow \uparrow }$ $=\sin (\theta /2)
T_{kRm^{\prime }}^{\ast },$ $(\Sigma ^{r})_{m^{\prime }k,\downarrow 
\downarrow }$ $=\cos (\theta /2)T_{kRm^{\prime }}^{\ast }$. From 
Eqs.(\ref{ae1}), (\ref{ae2}), and (\ref{ae3}), we obtain 

\begin{eqnarray}
G^{r}_{mn,\uparrow \uparrow } &=&(G_{0}^{r})_{mn,\uparrow \uparrow
}+\sum_{k,m'}(G_{0}^{r})_{mm^{\prime },\uparrow \uparrow }(\Sigma
_{R}^{r})_{m^{\prime }n^{\prime },\uparrow \uparrow }G^{r}_{n^{\prime
}n,\uparrow \uparrow } \nonumber \\
&+&\sum_{k,m'}(G_{0}^{r})_{mm^{\prime },\uparrow \uparrow }(\Sigma
_{R}^{r})_{m^{\prime }n^{\prime },\uparrow \downarrow }G^{r}_{n^{\prime
}n,\downarrow \uparrow }
\end{eqnarray}
with the corresponding self-energy functions $(\Sigma _{R}^{r})_{mn,\uparrow
\uparrow }$ \ and $(\Sigma _{R}^{r})_{mn,\uparrow \downarrow }$ given by

\begin{eqnarray}
(\Sigma_{R}^{r})_{mn,\uparrow \uparrow }&=&\sum_{k}T_{kRm}^{\ast
}T_{kRn}[g_{kR\uparrow }^{r}\cos ^{2}(\theta /2) \nonumber \\
&+&g_{kR\downarrow }^{r}\sin
^{2}(\theta /2)]
\label{aa1}
\end{eqnarray}

\begin{equation}
(\Sigma _{R}^{r})_{mn,\uparrow \downarrow }=\sum_{k}T_{kRm}^{\ast
}T_{kRn}[g_{kR\downarrow }^{r}-g_{kR\uparrow }^{r}]\sin (\theta /2)\cos
(\theta /2)
\label{aa2}
\end{equation}
Similarly, we can obtain other self-energy functions

\begin{equation}
(\Sigma _{R}^{r})_{mn,\downarrow \uparrow }=\sum_{k}T_{kRm}^{\ast
}T_{kRn}[g_{kR\downarrow }^{r}-g_{kR\uparrow }^{r}]\sin (\theta /2)\cos
(\theta /2)
\label{aa3}
\end{equation}

\begin{eqnarray}
(\Sigma _{R}^{r})_{mn,\downarrow \downarrow }&=&\sum_{k}T_{kRm}^{\ast
}T_{kRn}[g_{kR\uparrow }^{r}\sin ^{2}(\theta /2) \nonumber \\
&+&g_{kL\downarrow }^{r}\cos
^{2}(\theta /2)]
\label{aa4}
\end{eqnarray}
Eqs.(\ref{aa1})-(\ref{aa4}) are equivalent to Eq.(\ref{self}).

\section*{Figure Captions}

\begin{itemize}
\item[Fig. (1)]  (a). The current $I$ versus the voltage for different 
orientation angles: $\theta =0$ (solid line), $\theta =0.4\pi$ (dotted 
line), $\theta =0.8\pi$ (dashed line). Inset: the schematic plot showing
the TMR device considered in this work. (b). The current $I$ versus 
orientation $\theta$ at different bias $v=2,5,8,11,14$. Other parameters: 
$\Gamma_{\uparrow}=1.0$, $\Gamma_{\downarrow}=0.4$, $C_{1}=C_{2}=0.5$, 
$E_{F}-E_{0}=-2.$, and $\hbar=e=2m=1$.

\item[Fig. (2)] 
The current $I$ versus the orientation $\theta $ at
different temperatures: $\beta=1$ (solid line),  $\beta=0.5$ (dotted
line) $\beta=0.2$ (dashed line). Here the voltage is fixed at
$v=2$. The other parameters are the same as those of Fig.(1).

\item[Fig. (3)] The TMR ratio versus the voltage. Solid line $\beta 
=\infty $ and dotted line $\beta =1$. The other parameters are the same as
those of Fig.(1).

\item[Fig. (4)] The second order nonlinear conductance $G_{111}$, which is
the slope of conductance at zero bias, versus temperature parameter $\beta$.
Other parameters: $\Gamma_{1\uparrow}=1.0$, $\Gamma_{1\downarrow}=0.4$, 
$\Gamma_{2\uparrow}=0.8$, $\Gamma_{2\downarrow}=0.2$, $E_f=2.5$, and
$E_0=3$. 
\end{itemize}


\begin{thebibliography}{99}

\bibitem{prinz}
G.A. Prinz, Science, {\bf 282}, 1660 (1998).

\bibitem{baibich}
M. Baibich {\it et.al.}, Phys. Rev. Lett. {\bf 61}, 2472 (1988).

\bibitem{meservey}
For a review see, R. Meservey and P.M. Tedrow, Phys. Rep. {\bf 238}, 173 
(1994). 

\bibitem{monsma}
D.J. Monsma, J.C. Lodder, Th.J.A. Popma, and B. Dieny, Phys. Rev. Lett.
{\bf 74}, 5260 (1995); D.J. Monsma, R. Vlutters, J.C. Lodder, Science, {\bf
281}, 407 (1998).

\bibitem{egger}
S. Egger, C.H. Back, J. Krewer and D. Pescia, Phys. Rev. Lett. {\bf 83},
2833 (1999).

\bibitem{dax}
M. Dax, Semicond. Int. {\bf 20}, 84 (1997).
 
\bibitem{jull}
M. Julliere, Phys. Lett. A {\bf 54}, 225 (1975). 

\bibitem{moodera}
J.S. Moodera, L.R. Kinder, T.M. Wong, and R. Meservey, Phys. Rev. Lett.
{\bf 74}, 3273 (1995). 

\bibitem{slon}
J.C. Slonczewski, Phys. Rev. B {\bf 39}, 6995 (1989). 

\bibitem{moodera1}
J.S. Moodera, J. Nowak and R.J.M. van de Veerdonk, Phys. Rev. Lett. {\bf
80}, 2941 (1998).

\bibitem{zhang}
X. Zhang {\it et.al.}, Phys. Rev. B {\bf 56}, 5484 (1997).

\bibitem{bratkovsky}
A.M. Bratkovsky, Phys. Rev. B {\bf 56}, 2344 (1997).

\bibitem{fert}
J. Barna\'s and A. Fert, Phys. Rev. Lett. {\bf 80}, 1058 (1998).

\bibitem{levy}
S. Zhang, P.M. Levy, A.C. Marley and S.S.P. Parkin, Phys. Rev. Lett. {\bf
79}, 3744 (1997).

\bibitem{inoue}
J. Inoue and S. Maekawa, Phys. Rev. B {\bf 53}, R11927 (1996).

\bibitem{maclaren}
J.M. Maclaren, X.-G. Zhang and W. H. Butler, Phys. Rev. B {\bf 56}, 11827
(1997).

\bibitem{mathon}
J. Mathon, Phys. Rev. B {\bf 56}, 11810 (1997).

\bibitem{sheng}
L. Sheng, Y. Chen, H.Y. Teng, and C.S. Ting, Phys. Rev. B {\bf 59}, 480
(1999).

\bibitem{but1}
M. B\"uttiker, J. Phys. Condens. Matter {\bf 5}, 9361 (1993). 

\bibitem{bog}  
N.N. Bogoliubov, J. Phys. USSR, {\bf 11}, 23(1947).

\bibitem{jauho}  
A.P. Jauho, N.S. Wingreen, and Y. Meir, Phys. Rev. B {\bf 50},
5528(1994).

\bibitem{datta1}  S. Datta, {\it Electronic Transport in Mesoscopic 
Systems}, (Cambridge University Press, New York, 1995).

\bibitem{datta}  M. P. Anantram and S. Datta, Phys. Rev. B {\bf 51}, 7632
(1995).

\bibitem{stafford}  
C.A. Stafford, Phys. Rev. Lett. {\bf 77}, 2770 (1996).

\bibitem{jwang}  
J. Wang, et. al., Phys. Rev. Lett {\bf 80}, 4277 (1998).

\bibitem{christen}
T. Christen and M. B\"uttiker, Phys. Rev. Lett. {\bf 77}, 143 (1996). 

\bibitem{pedersen}
M. H. Pedersen and M. B\"uttiker, Phys. Rev. B {\bf 58}, 12993 (1998). 

\bibitem{ma1}
Z.S. Ma, J. Wang and H. Guo, Phys. Rev. B {\bf 57}, 9108 (1998); 
Phys. Rev. B {\bf 59}, 7575 (1999). 

\bibitem{wbg1}
B.G. Wang, J. Wang, and H. Guo, to appear in J. Appl. Phys. 1999.

\bibitem{foot}
Here we have used the Thomas-Fermi approximation.

\bibitem{but5}
T. Christen and M. B\"uttiker, Europhys. Lett. {\bf 35}, 523 (1996). 

\bibitem{caroli}  
C. Caroli, et.al., J. Phys. C {\bf 4}, 916 (1971); \ \ S. Hershfield,
J. H. Davies, and J. W. Wilkins, Phys. Rev. B {\bf 46}, 7046 (1992).

\end{thebibliography}
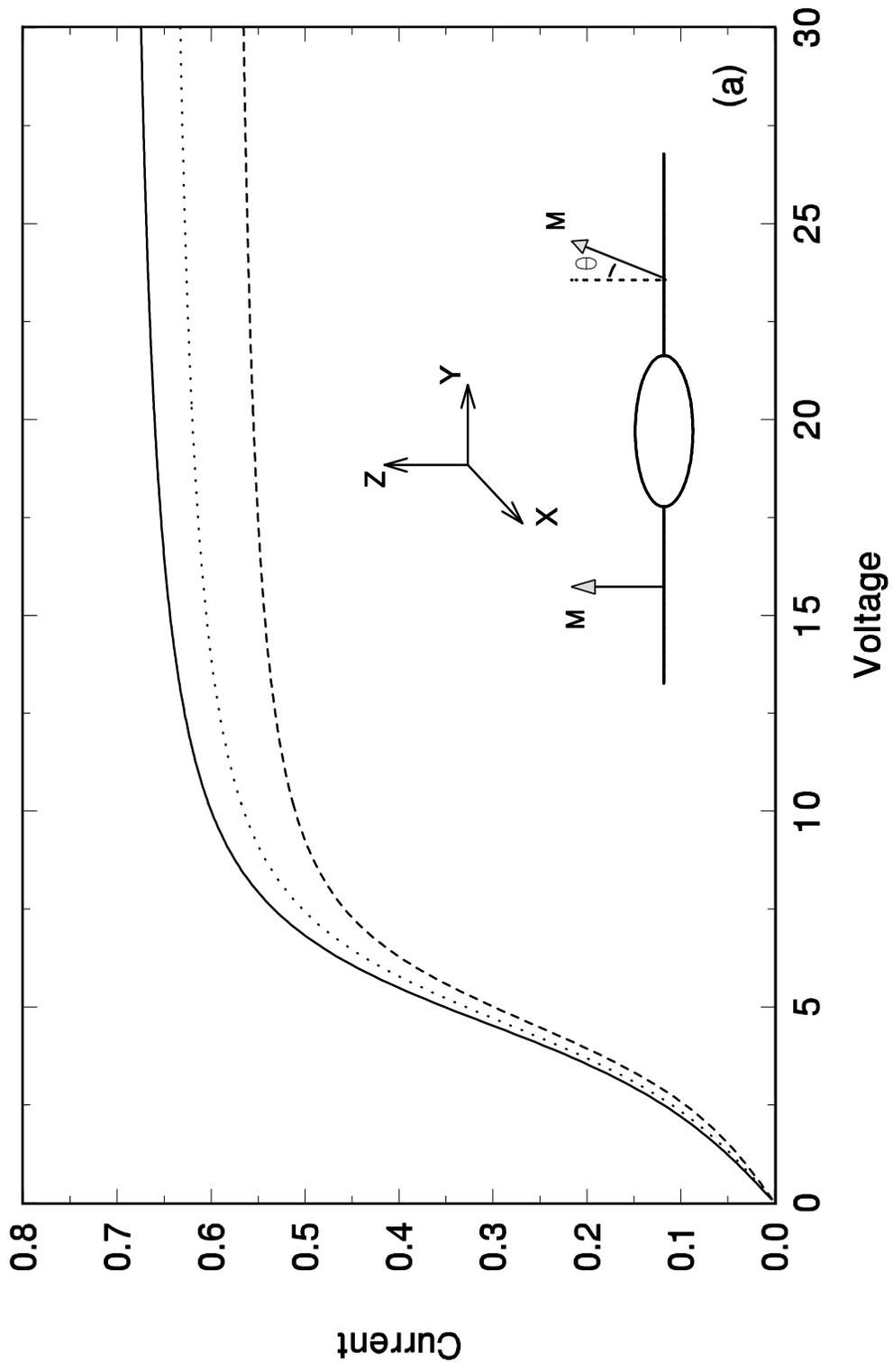
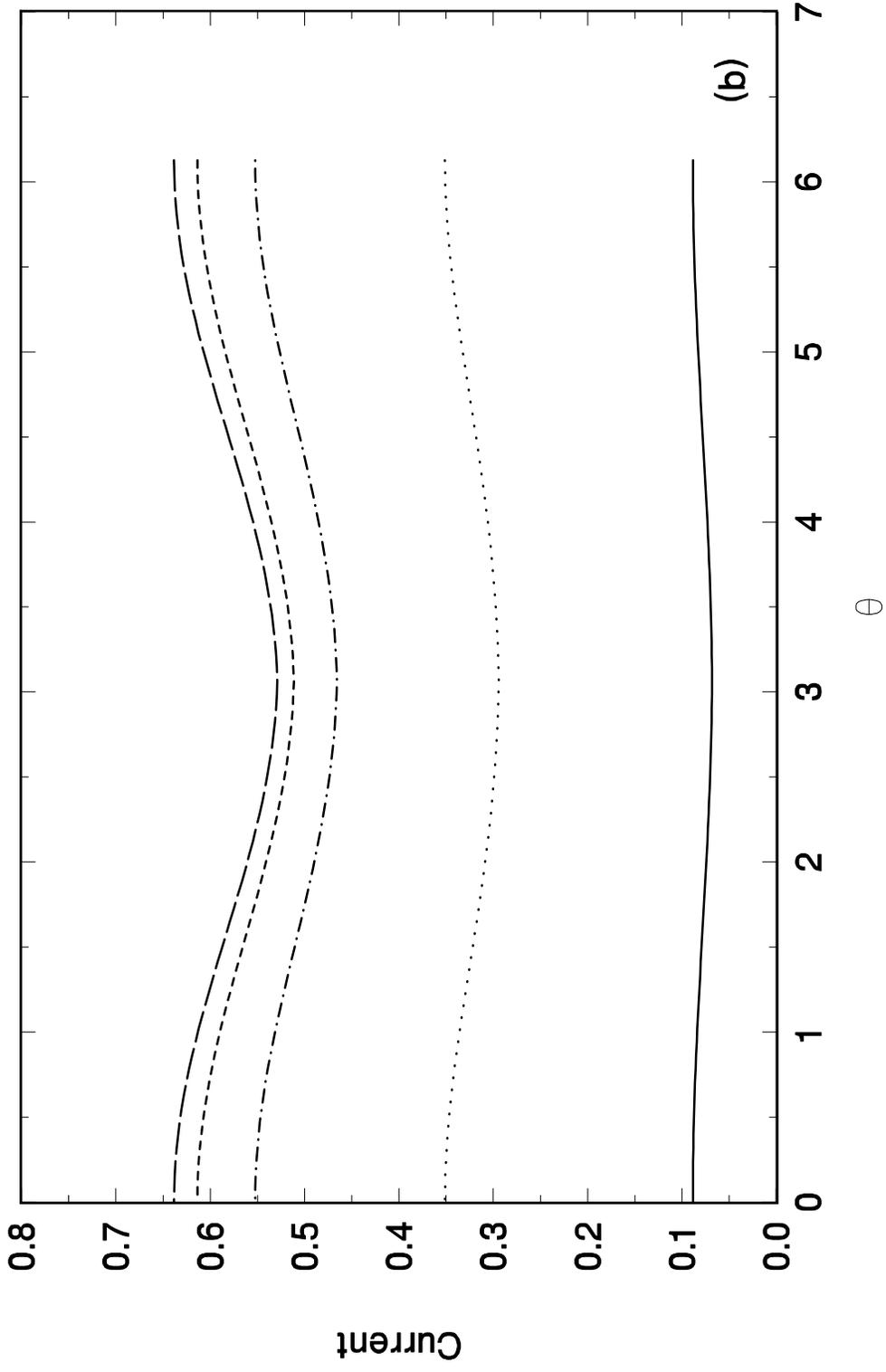
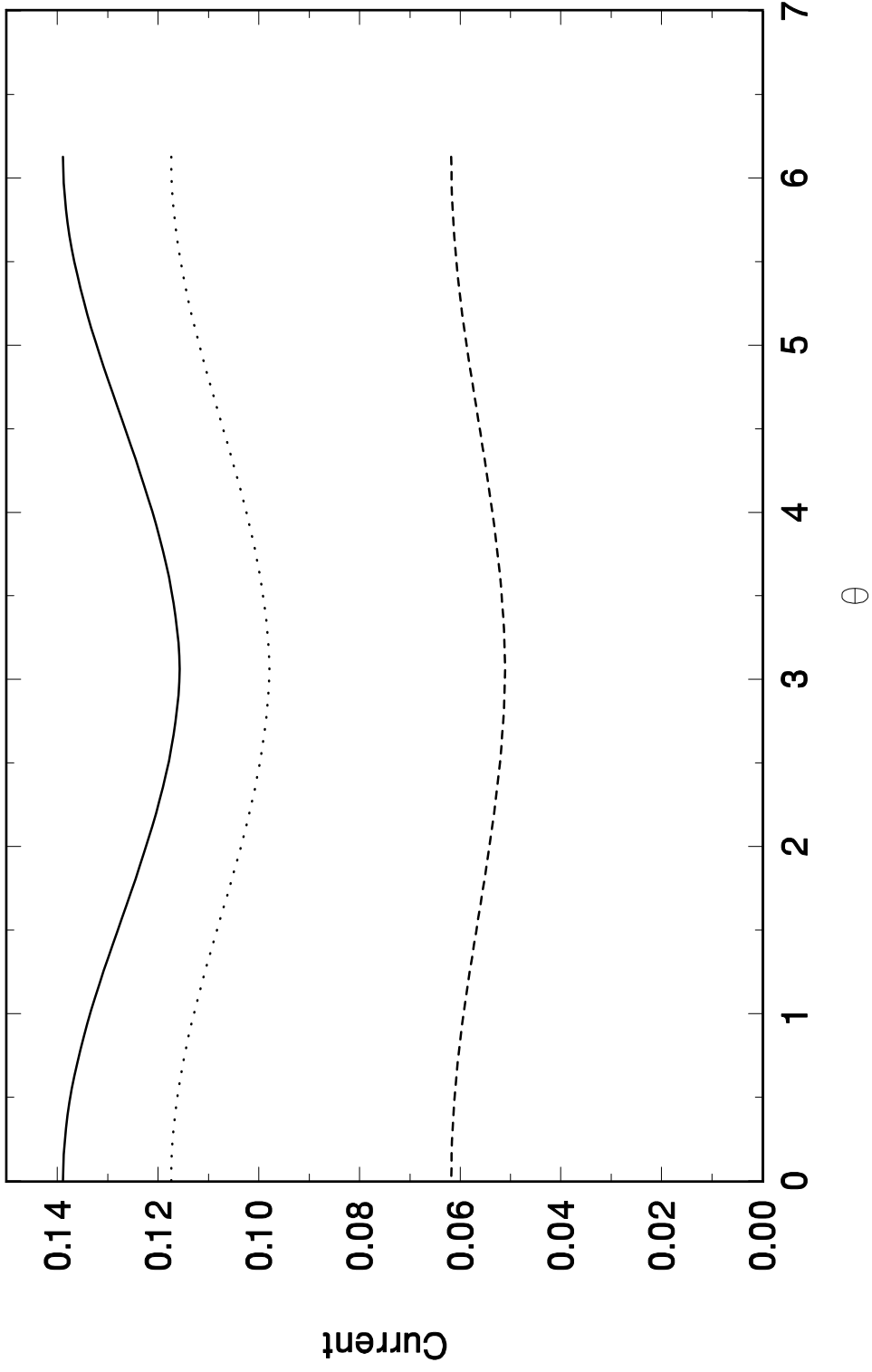
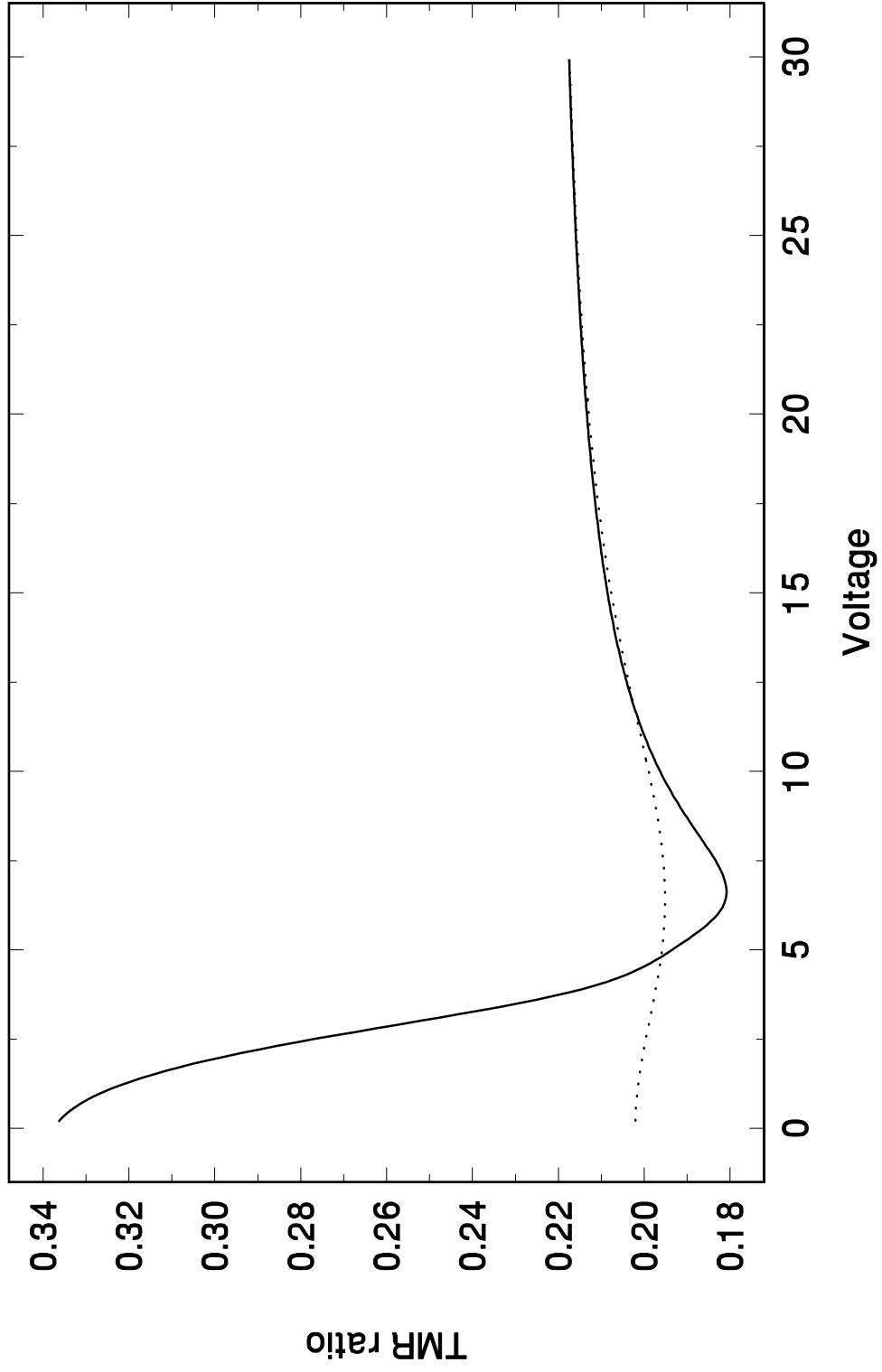
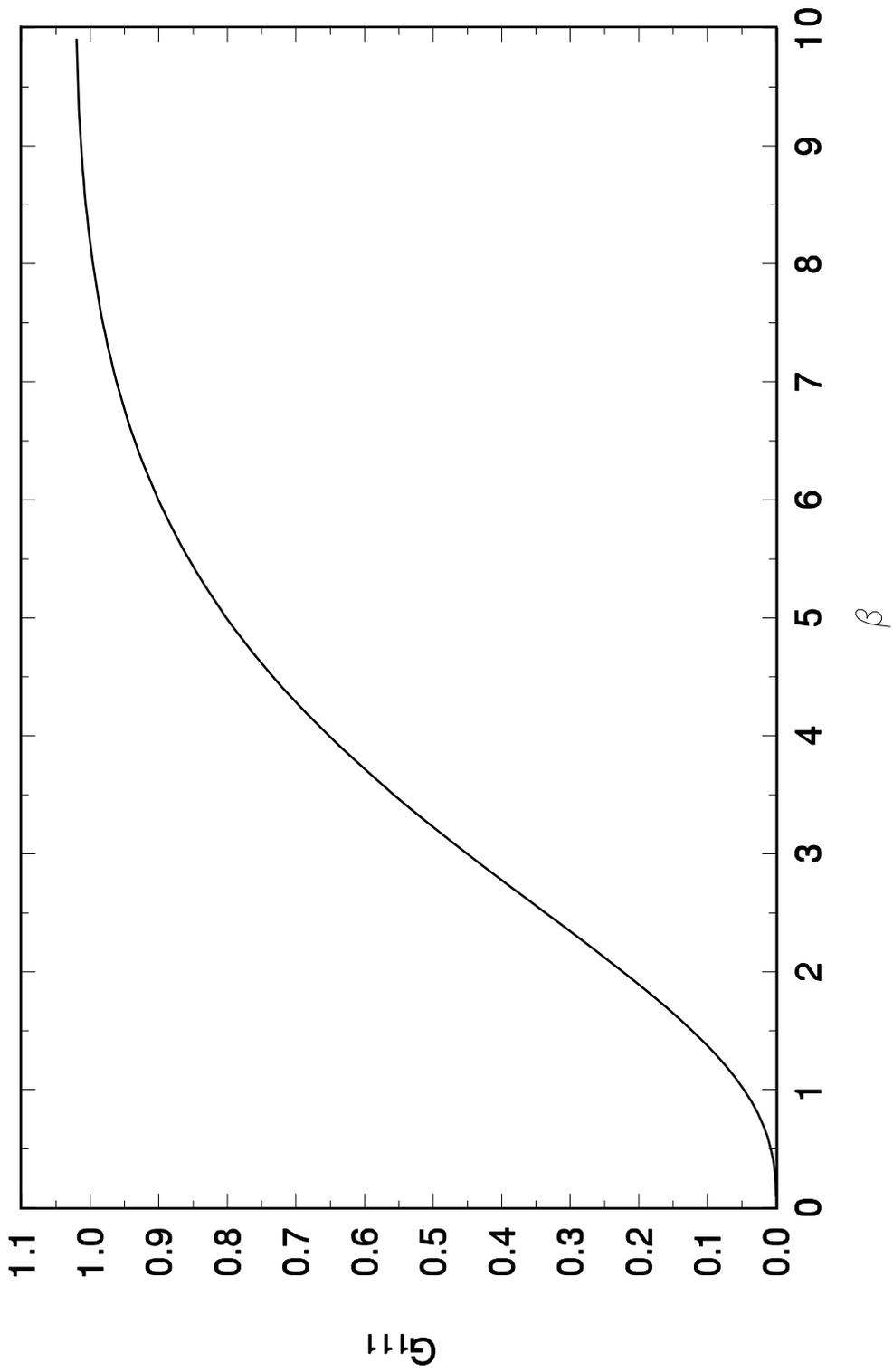
\end{document}